\documentclass[aps,pra,reprint,groupedaddress,amsmath]{revtex4-1}

\usepackage{graphicx}
\usepackage{dcolumn}
\usepackage{amsmath}
\usepackage{lipsum}
\usepackage{xcolor}
\usepackage{physics}

\begin{document}


\title{Proof-of-principle demonstration of vertical gravity gradient measurement using a single proof mass double-loop atom interferometer}


\author{I. Perrin}
\affiliation{DPHY,ONERA, Universit\'{e} Paris Saclay, F-91123 Palaiseau, France}

\author{M. Cadoret}
\email{malo.cadoret@lecnam.net}
\affiliation{LCM-CNAM, 61 rue du Landy, 93210, La Plaine Saint-Denis, France}

\author{Y. Bidel}
\affiliation{DPHY,ONERA, Universit\'{e} Paris Saclay, F-91123 Palaiseau, France}

\author{N. Zahzam}
\affiliation{DPHY,ONERA, Universit\'{e} Paris Saclay, F-91123 Palaiseau, France}

\author{C. Blanchard}
\affiliation{DPHY,ONERA, Universit\'{e} Paris Saclay, F-91123 Palaiseau, France}

\author{A. Bresson}
\affiliation{DPHY,ONERA, Universit\'{e} Paris Saclay, F-91123 Palaiseau, France}


\date{\today}

\begin{abstract} 
We demonstrate a proof-of-principle of direct Earth gravity gradient measurement with an atom interferometer-based gravity gradiometer using a single proof mass of cold $^{87}$Rb atoms. The atomic gradiometer is implemented in the so-called double-loop configuration, hence providing a direct gravity gradient dependent phase shift insensitive to DC acceleration and constant rotation rate. The atom interferometer (AI) can be either operated as a gravimeter or a gradiometer by simply adding an extra Raman $\pi$-pulse. We demonstrate gravity gradient measurements first using a vibration isolation platform and second without seismic isolation using the correlation between the AI signal and the vibration signal measured by an auxilliary classical accelerometer. The simplicity of the experimental setup (a single atomic source and unique detection) and the immunity of the AI to rotation-induced contrast loss, make it a good candidate for onboard gravity gradient measurements.

\end{abstract}


\maketitle



\section{Introduction}
Light-pulse Atom Interferometers (AIs) use short pulses of light to split, redirect, and then recombine cold atoms used as a matter-wave source. Since their advent in the 1990's \cite{Borde1989,Kasevich1991}, they have demonstrated to be extremely sensitive and accurate sensors very useful in fundamental physics research where they have been used to measure fundamental constants \cite{Muller2018,Bouchendira2011,Fixler2007,Rosi2014}, test the equivalence principle \cite{Fray2004,Bonnin2013,Schlippert2014,Tarallo2014,Zhou2015}, put bounds on theories of dark energy \cite{Hamilton2015}, probe quantum superposition at the macroscopic level \cite{Kovachy2015} as well as measuring gravito-inertial force such as  gravity acceleration  \cite{Peters2001,Gillot2014,Hu2013,Bidel2013,Hauth2013}, rotations \cite{Gustavson1997,Gauguet2009,Tackmann2012} and gravity gradient \cite{McGuirk2002,Sorrentino2014,Duan2014}. Most of these works consist in laboratory experiments but atom interferometer's inherent long term stability  and accuracy have led to a global push towards performing experiments outside laboratory environment \cite{Geiger2011,Bidel2013,Bidel2018}. Moreover, cold atom-based gravity sensors have started to be commercialized, hence targetting out of the lab applications. In this context, development of gravity gradiometers are also particularly attractive as they complement pure gravity measurements and find variety of applications including geodesy \cite{Carraz2014}, geophysics \cite{Nabighian2005} and inertial navigation \cite{Jekeli2005}. 
Whereas an atomic gravimeter sensitivity  is often limited by vertical vibration noise, it is not the case in a conventional atomic gradiometer where the gravity gradient is derived from the differential measurement of two simultaneous atom interferometers performed at two locations. However, this requires to use two clouds of cold atoms spatialy separated. This can for example be achieved by using laser-cooled atomic sources originating from two separate 3-dimensional magneto-optical traps (3D-MOT) \cite{McGuirk2002}, or  by launching the atoms from a single MOT using moving molasses \cite{Sorrentino2014} or Bloch oscillations as an atomic elevator \cite{Langlois2017,Cadoret2008} or using Large Momentum Transfer (LMT) beam splitters combining Bragg pulse and Bloch oscillations \cite{Esteythesis}. Although these techniques have proven to work in laboratory environment, their complexity could still be an issue regarding their implementation for onboard applications where simple and compact instruments are required.\newline
In this paper, we  perform a \emph{proof-of-principle} experimental demonstration of an alternative method consisting in a direct measurement of the vertical gravity gradient with only one source of cold 87 rubidium atoms in the presence of vibration noise. We use a  double-loop four-pulse AI geometry as proposed initially in \cite{Clauser1988} for gravity gradient measurements which was investigated in \cite{chungthesis} and now used in several experiments such as rotation rate measurements in atomic fountain configurations \cite{Stockton2011,Dutta2016} or  for low frequency vibration noise rejection in the context of airborne tests of the Weak Equivalence Principle (WEP) using atom interferometry\cite{Geiger2011}. We perform vertical gravity gradient measurement with and without a passive isolation vibration platform and show that in the presence of parasitic ground vibrations the correlation of the vibration signal measured by a classical accelerometer \cite{Lautier2014,Langlois2017} allow to recover the interference fringes and extract the vertical gravity gradient. Moreover, we make a study of the systematics when using this double-loop AI geometry. \newline
The paper is organized as follow: section \ref{Four pulse atom interferometer-based gravity gradiometer} presents the double loop four-pulse AI used to measure the vertical gravity gradient. Section \ref{setup} presents the experimental setup. Section \ref{GG1} describes the vertical gravity gradient measurement performing the AI with a passive vibration isolation platform and Section \ref{GG2} presents the measurement without seismic isolation, using the correlation technique. Section\ref{Systematic effects} presents a study of the major systematic effects which affect the measurement. Finally in Section \ref{Discussion}, a discusssion on  scale factor comparisons between dual cloud AI versus single cloud four-LPAI is made and possible improvements of the measurement are presented.

\section{Four pulse Atom interferometer-based gravity gradiometer}
\label{Four pulse atom interferometer-based gravity gradiometer}
\begin{figure}
\centerline{\includegraphics[width=8cm, scale=1.5]{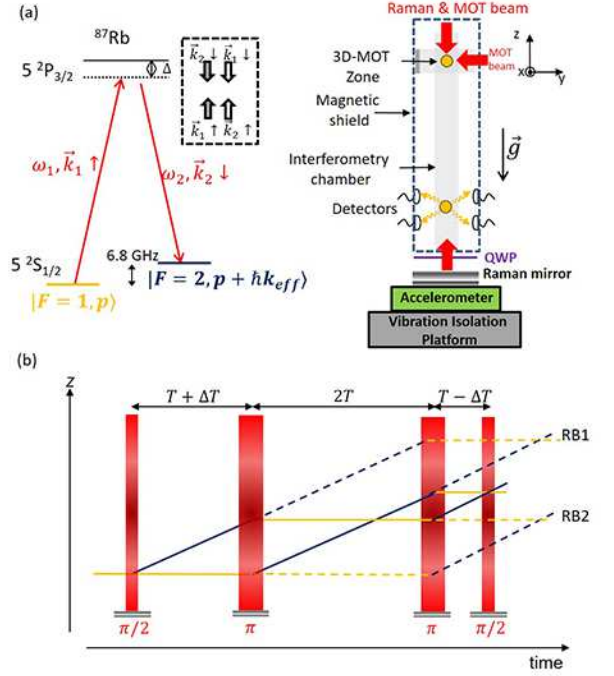}}
\caption
{ (a) Left: Energy level scheme of rubidium D2-line. Two-photon Raman transitions performed on a single cloud of 87 rubidium atoms in vertical configuration.($\Delta$ is the one photon detuning from the electronic transition.) Right: Schematic of the experimental setup. The Raman laser beams are aligned along gravitational acceleration $g$.PD: Photodiode.(b) Space-time recoil diagram in the absence of gravity of the four pulse double-loop AI based gravity gradiometer. A time asymmetry $\Delta T$ is implemented to suppress parasitic Ramsey-Bord\'{e} interferometers (labeled RB1 and RB2 on figure) (dash line) due to imperfect mirror pulses. }
\label{fig:figure1}
\end{figure}

We consider a four pulse AI. The matter-wave beam splitters and mirrors are based on two-photon counterpropagative Raman transitions between the $F=1$ and $F=2$ hyperfine ground states of rubidium 87 atoms (see FIG.\ref{fig:figure1}(a)). An atom initially in state $\ket{F=1,p} $ is coupled to state  $\ket{F=2,p+\hbar k_{\mathrm{eff}}}$ where $\hbar k_{\mathrm{eff}}$ is the two-photon momentum transfer. Here $\vec{k}_{\mathrm{eff}}=\vec{k}_1-\vec{k}_2$ is the effective wave vector (with $\vert\vert \vec{k}_{\mathrm{eff}}\vert\vert=\vert\vert \vec{k}_1\vert\vert+ \vert\vert\vec{k}_2\vert\vert$ for counterpropagative transitions). The AI consists of the light-pulse sequence $(\pi/2-\pi-\pi-\pi/2)$ which differs from a usual  Mach-Zehnder atomic gravimeter by the presence of an additional $\pi$-pulse and a different time sequence. Practically, a first $\pi/2$-pulse creates an equal superposition of ground ($F=1$) and excited ($F=2$) states. Then, two $\pi$-pulses redirect the two atomic paths letting the wave packets crossing each other in between, and a final $\pi/2$-pulse interferes the wavepackets. Thus, this sequence leads to a double-loop geometry. In our configuration, $\vec{k}_{\mathrm{eff}}$ is aligned with the local gravity acceleration $\vec{g}$, for all Raman pulses. This sequence allows to measure the time derivative of the acceleration of the free-falling atoms. 

Although a symmetric configuration is necessary to fully cancel the phase contribution due to constant acceleration \cite{Cadoret2016}, a time asymmetry $\Delta T$ is introduced to avoid interference of parasitic Ramsey-Bord\'{e} interferometers due to imperfect $\pi$ pulses \cite{Dutta2016}, see FIG\ref{fig:figure1}(b). In the absence of time asymmetry, these parasitic interferometers would close at the same time as the main interferometer and generate amplitude noise and a possible bias on the gravity gradient determination.\newline
In the short, intense-pulse limit, the phase shift along the Raman laser direction of propagation, is expressed as \cite{Cadoret2016}:
\begin{equation}
\begin{array}{lll}
\Delta \Phi=&4 (k_{\mathrm{eff}}g-\alpha)T\Delta T- (2 k_{\mathrm{eff}}v_z T^3+4k_{\mathrm{eff}}g T^4)\Gamma_{zz} \\
& +4k_{\mathrm{eff}}a_{\mathrm{vib}}T \Delta T-2k_{\mathrm{eff}}\dot{a}_{vib} T^3\\
&\equiv \varphi_{g}+ \varphi_{grad}+ \varphi_{vib}
\end{array}
\label{eq:phasegradio}
\end{equation}
where $g$ is the Earth gravity acceleration along the Raman laser beams, $\alpha$ the radio frequency chirp rate applied to the effective Raman frequency to compensate the Doppler shift induced by the atom free-fall in order to keep resonance, $\Gamma_{zz}$ the vertical gravity gradient component, $v_z$ the initial atomic velocity along the vertical $z$ axis at the first Raman pulse, $a_{vib}$  the mirror acceleration, $\dot{a}_{vib}$ its time derivative, and $T$ the time between the Raman $\pi/2$ and $\pi$ pulses in absence of timing asymmetry. In Eq.(\ref{eq:phasegradio}), the contribution to the phase shift contains three separate terms. The first term $\varphi_{g}$ is a remaining sensitivity to gravity acceleration induced by the timing assymetry $\Delta T$. We have embedded the laser phase $\alpha T\Delta T$ in this term. The second term is the gravity gradient dependent phase shift $\varphi_{grad}$, and finally the third term  is the phase shift induced by vibrations which we denote $\varphi_{vib}$. This vibrational phase noise  may prevent from discriminating spatial acceleration variations from time varying acceleration variations, and remains an issue for gravity gradient measurements performed in this double-loop geometry. Nevertheless, to circumvent this problem the AI can be operated  using a passive vibration isolation platform ($\varphi_{vib}\simeq 0$) or by estimating the $\varphi_{vib}$ phase term induced by the Raman mirror vibration using the acceleration noise recorded by an auxilliary classical accelerometer rigidly fixed to the Raman mirror. We have operated the AI using these two schemes.
For both schemes the atomic gradiometer is operated in its most sensitive configuration with  $T=38,6$ ms limited by the falling distance (see \ref{setup}), a timing asymmetry $\Delta T= 300\,\mu$s , an initial velocity at first Raman pulse $v_z=0.38$ m/s, leading to phase shift values reported in Table \ref{tab:tab1}. For our application we have neglected the effect of the recoil phase shift.

\begin{table}
\caption
{
\label{tab:tab1}
Phase shift terms numerical values assuming the following:$k_{\mathrm{eff}}=\frac{4\pi}{\lambda}=1.61 \times 10^{7}$
m$^{-1}$ with $\lambda=780$ nm the wavelength of the transition, vertical Earth gravity gradient  $\Gamma_{zz}=3.1\times 10^{-6}$ s$^{-2}$ (assuming a spherical symmetric Earth) with $g=9.81$ m.s$^{-2}$, interferometer pulse timing asymmetry of $\Delta T= 300\, \mu$s, $T= 38.6$ ms, initial velocity $v_z=0.38$ m.s$^{-1}$. The last term correponds to the recoil phase shift ($v_r=5.89$ mm.s$^{-1}$). The gravimeter phase shift term is taken as reference.}
\begin{ruledtabular}
\begin{tabular}{ccc}

			\rule[0cm]{0pt}{0.2cm} Phase term & Absolute numeric value  & Relative phase\\
			\rule[0cm]{0pt}{0.2cm}  & [mrad]  &\\
			\hline
						\rule[0.2cm]{0pt}{0.2cm} $4k_{\mathrm{eff}}gT\Delta T$ &$7.3\times 10^6 $ & $1$\\
\rule[0.2cm]{0pt}{0.2cm} $4k_{\mathrm{eff}}\Gamma_{zz}gT^4$ &4.3 &$ 6\times 10^{-7}$\\
			\rule[0.2cm]{0pt}{0.2cm} $2 k_{\mathrm{eff}}\Gamma_{zz}v_z T^3$ &2.2& $3\times 10^{-7}$ \\
			$k_{\mathrm{eff}}\Gamma_{zz}v_{r}T^3$&$1.7\times 10^{-2}$&$2\times 10^{-9}$\\

\end{tabular}
\end{ruledtabular}
\end{table}

\section{Experimental setup}

\label{setup}
In this section we describe the main parts of the apparatus as well as the time sequence of the experiment. 
\subsection{Apparatus overview}
Our experimental setup is schematically shown in FIG.\ref{fig:figure1}.
It consists of a titanium vacuum chamber where the atoms are produced and interrogated and to which are connected ion pumps, getters and rubidium dispensers. The vacuum chamber is magnetically shielded with two cylindrical layers of $\mu$-metal. The cold atom source is produced at the top of the chamber using a 3D MOT configuration. The falling distance available for interferometry is 20 cm from the MOT. The two counter-propagating Raman beams are obtained with a phase-modulated laser at 6.8 GHz retro-reflected on a mirror (Raman mirror). The Raman laser beams enter the vacuum chamber through the top window. After passing through a quarter wave plate they are retroreflected by the Raman  mirror at the bottom of the setup, outside the vacuum, in order to realize the counterpropagating configuration, thus obtaining a lin $\perp$ lin configuration in the AI region. In this configuration, two pairs of counterpropagating Raman beams ($\uparrow \vec{k}_1 ,\downarrow \vec{k}_2$ and $\downarrow \vec{k}_1 ,\uparrow \vec{k}_2$ ) in the vertical direction are present. Degeneracy between the two pairs of Raman beams is lifted through Doppler shift induced by gravity during free fall of the atoms. In this configuration, only the Raman mirror needs to be isolated from ground vibrations. In our setup the mirror is rigidly linked to a classical accelerometer (Titan Nanometrics) the whole being fixed to a passive vibration isolation platform (Minus-K).

\subsection{Optical setup}
The laser system used for cooling, detecting and driving the interferometer pulses is similar to the one descibed in \cite{Theron2015}. Basically, it consists in a compact and robust laser system based on a single narrow linewidth Erbium doped fiber laser  at 1.5 $\mu$m, amplified in a 5 W Erbium doped fiber amplifier (EDFA) and then frequency doubled in a periodically poled lithium niobate (PPLN) crystal. A power of 450 mW is available at 780 nm. The Raman laser and the repumper are generated thanks to a fiber phase modulator at 1.5 $\mu$m allowing to be free from any phase lock loop between the two Raman lines.
\subsection{Experimental sequence}
\label{Experimental sequence}
The experimental sequence of the atomic gradiometer is the following:
first, a cold  $^{87}$Rb sample is produced in a 3-dimensional  MOT, loaded from a background vapor pressure of  $\sim 10^{-8}$ mbar. After 700 ms of trap loading, a stage of optical molasses and a microwave selection, we assemble $N_{at}\sim 5\times 10^{7}$ atoms in the magnetic insensitive groundstate $\ket{F=1, m_F=0}$ at a temperature $\Theta=3\,\mu$K. A push beam gets rid of the atoms left in state $F=2$. Then, after $38.6$ ms of free-fall we apply the AI sequence consisting in four Raman laser pulses of 8,16,16 and 8 $\mu$s. This time delay before the first Raman pulse is necessary to first lift degeneracy between the two pairs of Raman beams and second to minimize the impact of parasitic Raman lines (see section \ref{Systematic effects}). During the AI operation a bias magnetic field of 100 mG is applied. We set $T=38.6$ ms corresponding to total interrogation time $4T=154.4$ ms. 
Following the interferometer sequence we measure the proportion of atoms in the two output ports $F=2$ and $F=1$  of the interferometer using state selective vertical light-induced fluorescence detection. The fluorescence is collected thanks to collimation lenses and photodiodes in the perpendicular direction.
The measurement of the proportion of atoms $P$ in the state $F=2$ at the exit of the interferometer   is a sinusoidal function of the interferometric phase shift:
\begin{equation}
P=P_m +\frac{C}{2}\cos(\Delta\Phi)
\label{eq:proba}
\end{equation}
where $P_m$ is the fringe offset, and $C$ the fringe contrast which is in our case $C=0.1$. Interferometric fringes are thus obtained through scanning the interferometric phase. In our experiment we operate the interferometer with and without vibration isolation platform. Therefore, a scanning of the phase is obtained first  by varying the frequency chirp rate of the Raman laser and second by letting vibration noise operate a random sampling of the interferometric phase.\newline
The repetition rate of the experimental sequence is 1 Hz, including atom loading, state preparation, atom interferometry, and state detection. The whole experimental sequence timing and data aquisition is computer controlled.

\section{Gravity gradient measurement with vibration isolation}
\label{GG1}
In this section we present the vertical gravity gradient measurement when operating the AI with a vibration isolation platform.

\subsection{Measurement method}
\label{Measurement method1}
To measure the vertical gravity gradient we take advantage of the acceleration sensitivity induced by the timing assymetry $\Delta T$ of the interferometer. In presence of a vibration isolation platform ($\varphi_{vib}\approx 0$), interferometer fringes can be obtained by scanning the frequency chirp $\alpha$. The interference phase is obtained using the Fringe-Locking Method (FLM) similar to the one described in \cite{Bidel2013}, which determines the frequency chirp nulling the phase. The sign of the radio-frequency chirp is changed every two drops, hence reversing the sign of $\vec{k}_{\mathrm{eff}}$  to cancel some systematic effects.
Nevertheless, in our protocol, we supress sensitivity to gravity acceleration by periodically reversing the sign of $\Delta T$ (every 5 minutes), hence reversing the sign of the acceleration phase shift in Eq.(\ref{eq:phasegradio}). Nulling the phase shift in both configuration and taking the mean value leads to:
\begin{equation}
\Gamma_{zz}=\frac{2T\Delta T(\alpha^{0}_--\alpha^{0}_+)}{2k_{\mathrm{eff}}v_zT^3+4k_{\mathrm{eff}}gT^4}
\label{Eq:gammazz}
\end{equation}
where $\alpha^{0}_{+}$,($\alpha^{0}_{-}$) is the frequency chirp which nulls the phase for $\Delta T$ ($-\Delta T$) respectively.

\subsection{Gradiometer sensitivity}
\begin{figure}[h]
\centerline{\includegraphics[width=8cm]{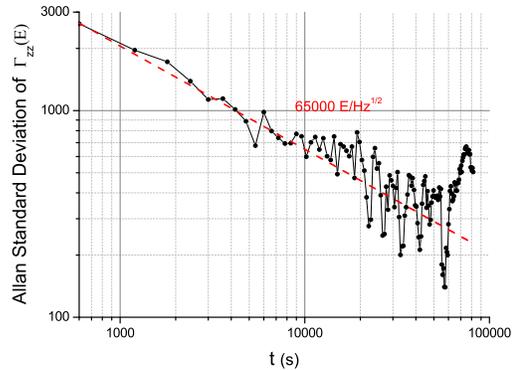}}
\caption
{ Allan standard deviation of the gravity gradient measurements. The dash line illustrates the $t^{-1/2}$ scaling. }
\label{fig:figure2}
\end{figure}

We have operated the gravity gradiometer continuously during 2 days using the experimental sequence decribed in (\ref{Experimental sequence}). The time assymetry $\Delta T$ is changed every 5 minutes. We obtained from equation $(\ref{Eq:gammazz})$ the uncorrected vertical gravity gradient mean value $\Gamma_{zz}=7600$ E. 
The Allan Standard Deviation (ADEV) on the gravity gradient measurements is shown on FIG.\ref{fig:figure2}.
Each point correponds to the measurements averaged over 10 minutes corresponding to the time necessary to operate th AI in two configurations ($+\Delta T$ and $-\Delta T$).
A short term sensitivity of 65000 E/$\sqrt{Hz}$ is obtained during the two days of measurements.
The stability of the gravity gradient measurement improves as $t^{-1/2}$ (where $t$ is the measurement time) and reaches  766 E after 2 hours. The sensitivity of our measurement is not limited by the contribution of residual vibration noise which has been measured with our low noise accelerometer at the level of 10000 E$/\sqrt{\mathrm{Hz}}$, but more by technical noise which we did not investigate further for this experimental demonstration.
We have investigated the main systematic effects which induce a bias on the gravity gradient measured value using this method. These systematics are presented in section \ref{Systematic effects}.

\section{Gravity gradient measurement using the correlation technique}

\label{GG2}
In this section we present gravity gradient measurement in the presence of vertical vibration noise.  

\subsection{Measurement method}
\begin{figure}
\centerline{\includegraphics[width=8cm]{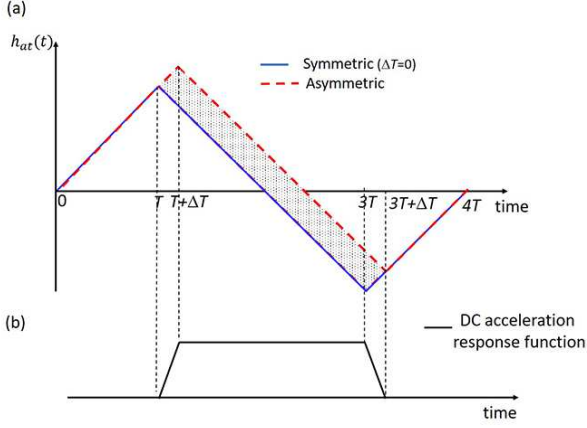}}
\caption{(a)Four pulse AI response function $h_{at}$  neglecting pulse duration for a symmetric (solid line) and asymmetric (dash line) configuration. (b) Response function of the resisual sensitivity to DC acceleration is obtained from the difference between asymmetric and symmetric four pulse AI response functions.}
\label{fig:figure3}
\end{figure}

First the  vibration isolation platform on which is fixed the Raman mirror is made non-floating. In the absence of vibration isolation $\varphi_{vib}\neq 0$ in Eq.\ref{eq:phasegradio}, thus the conventional FLM used in \ref{Measurement method1} is not applicable anymore as it requires phase fluctuations to be smaller than $\pi$. To circumvent the presence of vibration excess noise which washes out fringe visibility, we perform a correlation-based-technique \cite{Merlet2009,Barrett2015} combining the simultaneous measurements of the output signal $P$ of our interferometer and the one from a classical accelerometer fixed to the Raman mirror (see FIG(\ref{fig:figure1})).
The method is the following:\newline
First we held the laser phase fixed by setting the radiofrequency chirp $\alpha_0$ to its value compensating for gravity acceleration leading to $\varphi_{g}= 0$. This value of $\alpha_0$ is determined by operating the interferometer as a conventional 3-pulse Mach-Zehnder interferometer using $T=81.9$ ms (Where $T$ is the time between two-consecutive Raman pulses). Second, the AI is operated with the same experimental sequence except that the radiofrequency chirp sign is changed every measurement cycle. The atomic fringes are scanned due to random vibration noise. 
The probability $P$ of the interferometer is plotted versus the estimated induced vibration-phase $\varphi_{vib}^{E}$, which is numerically calculated at each cycle by convoluting the mirror acceleration $a_M(t)$ measured by the classical accelerometer, with the time response funtion $h_{at}(t)$ of the AI:
\begin{equation}
\varphi_{\mathrm{vib}}^{E}=k_{\mathrm{eff}}\int_{}^{}a_M(t)h_{at}(t)dt
\end{equation}
where $h_{at}$ is a double triangle-like function represented on FIG.(\ref{fig:figure3}) defined as:
\begin{equation}
h_{at}(t)=
\left\lbrace
\begin{array}{ccc}
\frac{t}{T^2}  & \mbox{if} & 0<t<T+\Delta T\\
\frac{2(T+\Delta T)-t}{T^2} & \mbox{if} & T+\Delta T<t<3T+\Delta T\\
\frac{t-4T}{T^2}& \mbox{if} & 3T+\Delta T<t<4T \\
0&&\mathrm{Otherwise.}
\end{array}\right.
\end{equation}

Finally, we perform a sinusoidal least-square fit of the data using the function:
\begin{equation}
P=A+\frac{B}{2}\cos(\varphi_{vib}^{E}+\delta \phi)
\label{eq:fit}
\end{equation}
where $A,B$ and $\delta\phi$ are free-parameters. 
Performing a measurement of the transition probability in four configurations ($\vec{k}_{\mathrm{eff}} \uparrow,\downarrow,\pm\Delta T$) where reversing the sign of $\vec{k}_{\mathrm{eff}}$ (e.g changing the sign of $\alpha$) allows to reject some systematics, and reversing the sign of $\Delta T$ suppresses residual dependence to constant acceleration, one can obtain the gravity gradient.

\subsection{Experimental results}

\begin{figure}
\centerline{\includegraphics[width=9cm]{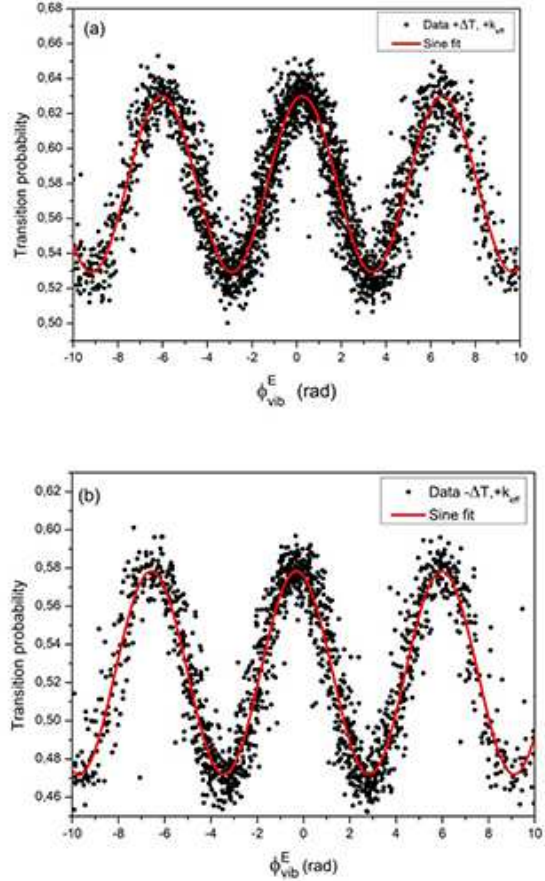}}
\caption
{Measured transition probabilities versus estimated vibration phase calculated from the signal of the classical accelerometer. The atomic gradiometer operates with total interrogation time 4T=154,4 ms. The solid line is a sinusoidal least-squares fit using equation (\ref{eq:fit}). (a) Interference fringes when using $+\Delta T$ timing assymetry.(b) Interference fringes obtained in the $-\Delta T$ configuration. Two other spectra are obtained reversing the sign of the wavevector.}
\label{fig:figure4}
\end{figure}

We performed gravity gradient measurement during two hours integration time. For the measurement, the sign of $\Delta T$ was changed after one hour integration time whereas  the direction of the effective wavevector was reversed every measurement cycle. Atomic fringes in  configuration ($\pm \Delta T$) are displayed on  FIG.\ref{fig:figure4} when operating the interferometer with total interrogation time $4T=154.4$ ms. 
We obtained 4 spectra in total corresponding to approximately 1800 points per fringe pattern. $\delta\varphi$ is estimated from the fit to the data points for each fringe pattern.  The gravity gradient is extracted from the phase offset $\delta\varphi$ considering the four  configurations ($\vec{k}_{\mathrm{eff}} \uparrow,\downarrow,\pm\Delta T$) leading to 4  phases ($\delta\varphi_{\uparrow,+},\delta\varphi_{\uparrow,-},\delta\varphi_{\downarrow,+},\delta\varphi_{\downarrow,-}$ ). Considering the experimental protocol one obtains:
\begin{equation}
\Gamma_{zz}=\frac{1}{4}\frac{\left(\delta\varphi_{\uparrow,+}+\delta\varphi_{\uparrow,-}+\delta\varphi_{\downarrow,+}+\delta\varphi_{\downarrow,-}\right)}{2k_{\mathrm{eff}}v_z T^3 + 4k_{\mathrm{eff}}gT^4}
\end{equation}  
corresponding to an uncorrected gravity gradient of $\Gamma_{zz}=3691$ E.
The sensitivity of the gradiometer is evaluated from the combined statistical uncertainty from the fitted fringes leading to $\delta\Gamma_{zz}=2355$ after 2 hours integration time.
The sensitivity of the measurement using the correlation technique is degraded by a factor of 3 in comparison with the measurement performed in presence of vibration isolation. First, a decrease by a factor of $\sqrt{2}$ may originate from the use of a Fringe Scanning (FS) method instead of a more sensitive Fringe Locking Method \cite{Hu2018}. 
Second, degradation of the sensitivity may come from non-perfect correlations due to several factors that we did not have time to investigate such as : misalignment between the classical accelerometer and the AI, unprecise knowledge of the mechanical accelerometer scale factor, uncertainty in the accelerometer transfer function, bias drift, among others. We estimate the classical accelerometer self-noise to limit our sensitivity at the level of 2 E$/\sqrt{Hz}$. An improvement of the sensitivity on our measurement is therefore possible. We give in section \ref{Discussion} some possible improvements of the method.

Systematic effects are studied in the next section.

\section{Systematic effects}
\label{Systematic effects}
In this section we present a study of the main systematics limiting the measurement of the gravity gradient and their related uncertainties.
\subsection{Effect of a slope on fringe offset}
In our experiment, we noticed that the fringes obtained by scanning $\alpha$ have a slope on the offset.  This slope is due to  a change in the resonance condition which appears because of a relatively large fringe spacing ($\propto 1/4T\Delta T$) relative to the fringe envelope ($\propto 1/\tau_{\pi/2}$). This slope is different for each of the four configurations and therefore when the FLM \cite{Bidel2013} is used, it is responsible of a bias on the gravity gradient measurement equal to:
\begin{equation}
\Delta\Gamma=\frac{\pi}{24 C k_{\mathrm{eff}}g T^5 \Delta T}\times A
\end{equation}
where $A$ is the slope of the fringe offset defined as:
\begin{equation}
P_m=P_{m0}+(\alpha-\alpha_0)A
\end{equation}
$A$ has been measured for the four configurations of the experiment ($\vec{k}_{\mathrm{eff}} \uparrow,\downarrow,\pm\Delta T$). From these four measurements one can obtain the value of $A$ to estimate the bias:
\begin{equation}
A_{\uparrow,\downarrow,\pm}=\frac{(A_{\uparrow,-}-A_{\downarrow,-})-(A_{\uparrow,+}-A_{\downarrow,+})}{2}
\label{eq:A}
\end{equation}
Using the value of $A$ from equation $(\ref{eq:A})$, one obtains an estimated bias equal to $4351$ E $\pm\, 430$ E in our configuration.

\subsection{Raman detuning }
In our experiment, the one photon light-shift is largely canceled by  adjusting the intensity ratio between the Raman lasers and  using the effective wave vector reversal protocol. Nevertheless,  sensitivity to laser detuning remains through two-photon light-shift (TPLS)\cite{Clade2006} and initial velocity of the atoms.
\subsubsection{Residual sensitivity to atom velocity}
\begin{figure}
\centerline{\includegraphics[width=8cm]{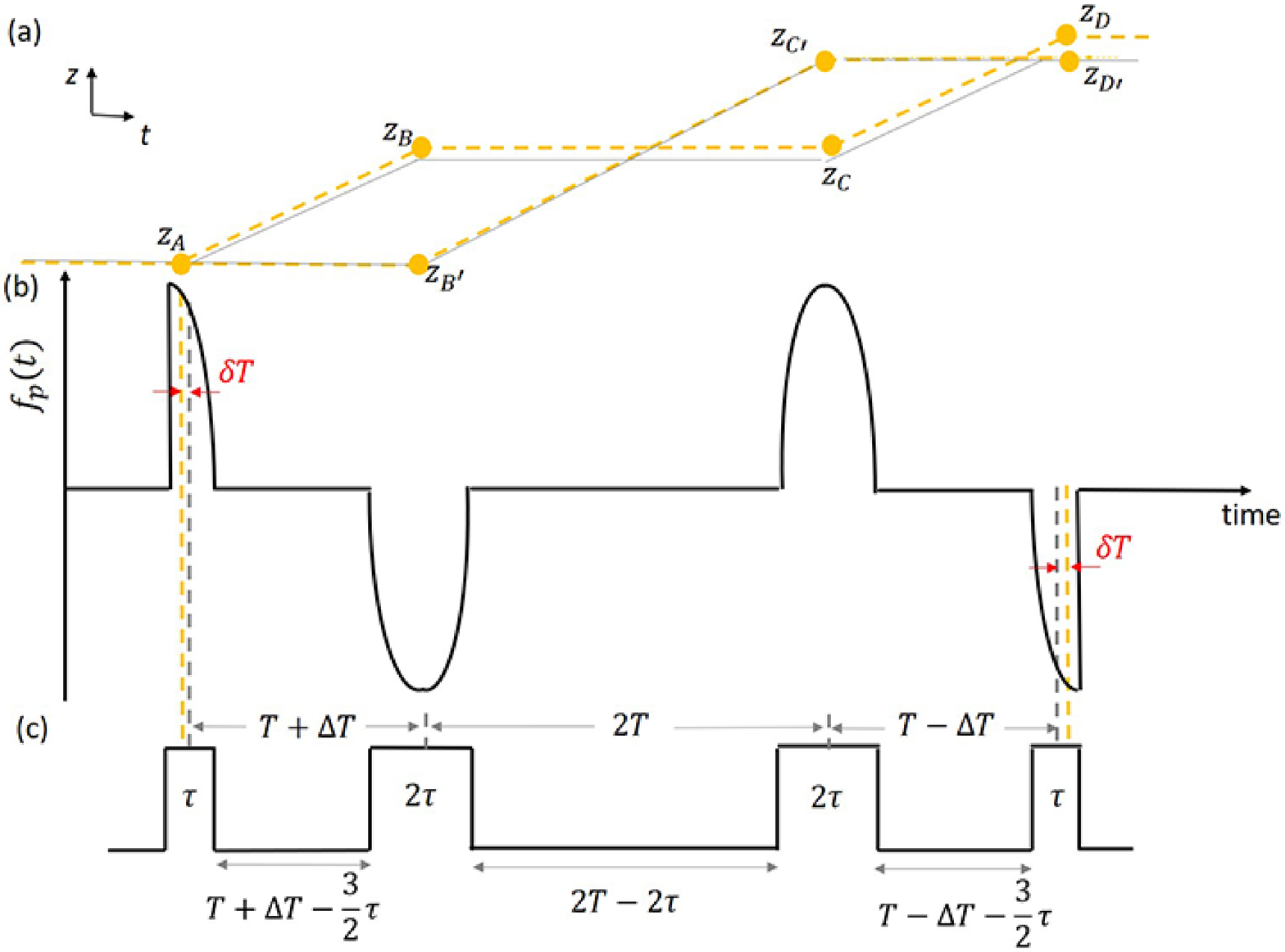}}
\caption
{(a): Space-time diagram of the four pulse AI in absence of gravity considering the photon momentum exchange to be centered on the pulse (straight line) and taking into consideration the response function of the interferometer during the pulses (dash-line). The dissymetry due to the presence of the extra $\pi$-pulse with respect to a Mach-Zehnder-type AI, leads to a separation phase shift at the exit of the AI ($z_D\neq z_D'$).(b) Sensitivity function to displacement $f_{p}(t)$ considering finite Raman pulses \cite{Bonninthesis}.
(c)  Timing diagram of the four pulse AI defining time with respect to the center of the Raman light-pulse.}. 
\label{fig:figure5}
\end{figure}

\begin{figure}
\centerline{\includegraphics[width=8cm]{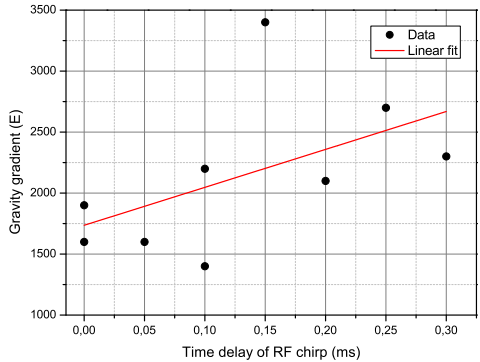}}
\caption
{Gravity gradient measurement as a function of the time delay to switch on the radiofrequency chirp. The slope given by the linear fit is $\approx 3000$ E/ms.}
\label{fig:figure6}
\end{figure}

Contrary to a three-pulse Mach-Zehnder-type AI, a four pulse AI exhibits a phase sensitivity to initial velocity  due to a dissymetry between the first and last $\pi/2$-pulses induced by the presence of the extra $\pi$-pulse. This dissymetry appears when considering finite Raman pulses and is reponsible of a phase shift separation at the exit of the interferometer. This effect is  illustrated on (FIG.\ref{fig:figure5}) which represents the atomic trajectories (FIG.\ref{fig:figure5}(a)) as a function of the position response function of the AI (FIG.\ref{fig:figure5}(b)).
To circumvent this effect, one has to precisely adjust the timing $\delta T$ between the two $\pi$-pulses and between the $\pi/2$ and $\pi$-pulses in order to have a closed interferometer and to have no dependency on the atom velocity on the phase shift.\newline
Assuming perfect $\pi/2$-pulses, the time separation between the center of each pulses should be equal to:
\begin{eqnarray}
T_{\frac{\pi}{2}-\pi}&=&T+\Delta T+\delta T\\ \nonumber
T_{\pi-\pi}&=&2T\\
T_{\pi-\frac{\pi}{2}}&=&T-\Delta T+\delta T\nonumber
\end{eqnarray}
where $\delta T=\left(\frac{2}{\pi}-\frac{1}{2}\right)\tau_{\frac{\pi}{2}}$.\newline
Thus, a time compensation $\delta T=1\,\mu$s is added to the sequence to compensate for this effect.
Nevertheless, experimental defaults such as non perfect $\frac{\pi}{2}$ pulse or pulse shape asymmetry could affect this timing. We thus measured experimentally the phase shift versus the atom velocity at the first Raman pulse (see FIG.\ref{fig:figure6}). This was done by varying the time delay at which  the radiofrequency chirp $\alpha$ used to compensate the Doppler shift is switched on. One obtains a slope of 3000 E/ms. The uncertainty on the atom velocity is estimated at 1 mm/s leading to an uncertainty on the gravity gradient equal to 300 E.
\subsubsection{Two-photon light shift}
In our setup there are two pairs of beams which can drive the counterpropagative Raman transition (see FIG.\ref{fig:figure1}) with wavevector $\pm \vec{k}_{eff}$. Consequently, the pair which is out of resonance will induce a two-photon light shift (TPLS).

\begin{figure}
\centerline{\includegraphics[width=8cm]{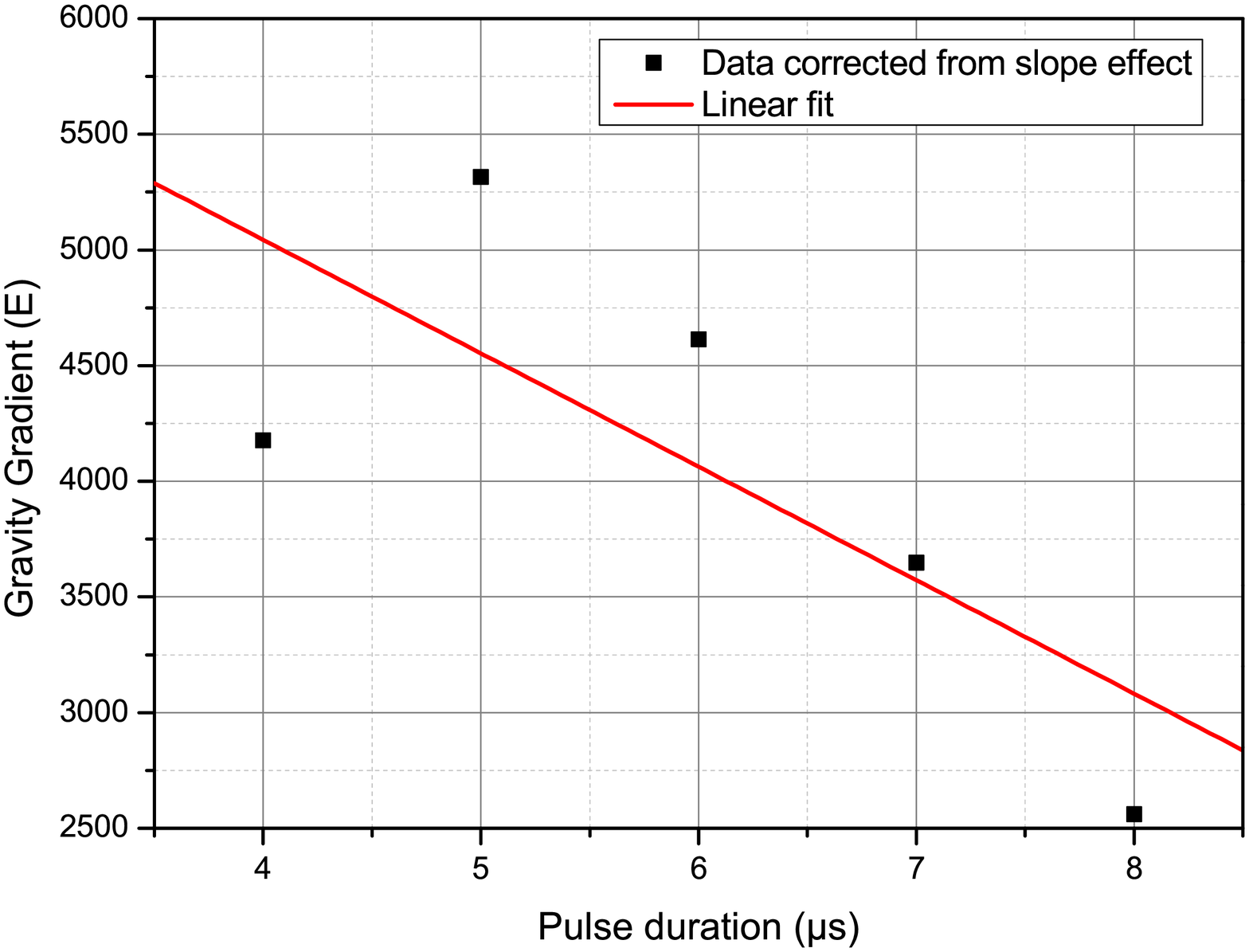}}
\caption
{ Variation of the gravity gradient value due to two-photon light shift (TPLS) versus $\pi/2$-pulse duration keeping Rabi frequency $\Omega_{eff}$ constant. The Raman $\pi$-pulses are kept constant to preserve the interferometer's contrast. All data points are corrected from the effect of the slope on fringe offset. The slope given by the fit is $-490\,E/\mu$s.}
\label{fig:figure7}
\end{figure}

The TPLS is estimated by measuring the gravity gradient as a function of the $\pi/2$-pulse duration.For the measurements, the mirror pulses are kept constant to preserve the interferometer's contrast. Moreover, as the contrast changes with the pulse duration, the measurements are corrected from the slope effect (see FIG. \ref{fig:figure7}).
From the linear fit of the measurements we obtained a bias of $\Delta \Gamma=- 4351$ E $\pm 2019$ E (uncertainty from the fit).

\subsection{Additional Raman laser lines}
Our method of generating the Raman laser by modulation leads to the presence of additional laser lines inducing a supplementary phase shift which if not corrected, induces an error on the gravity gradient measurement. We can numerically calculate this supplementary phase shift according to \cite{Carraz2012} and transposing to the case of a four-pulse AI. In the symmetric configuration of the interferometer ($\Delta T=0$) it is possible to find an AI configuration where this phase shift is equal to zero. This corresponds to the case where the distance between the position of the atoms at the moment of the four Raman pulses are multiples of the microwave wavelength $\Lambda= c/2\omega_{HFS}$. According to \cite{Carraz2012} and transposing to the four pulse AI the conditions on interrogation time $T$ and initial velocity of the atoms are  given by: 
\begin{equation}
T=\sqrt{\frac{n\Lambda}{3g}}\,\,\, \mathrm{and}\,\,\, v_{z}=\frac{n'\Lambda - \frac{1}{2}gT^2}{T}\,\,\mathrm{with}\,\,\, n,n' \in N
\end{equation}
Considering the available free-fall height of 20 cm one finds $n=2$ and $n'=1$ leading to
$T=38.6$ ms and initial atomic velocity at first Raman pulse  $v_z=gT=0.38$ m.s$^{-1}$.
However, timing assymetry $\Delta T=300\,\mu$s required to avoid the presence of extra  RB interferometers, changes the atoms position and prevents from operating the interferometer in this optimal configuration. We have numerically calculated the error on the gravity gradient for time assymetry $\Delta T=300\,\mu$s. This error is a periodic function of the atom-mirror position $z_M$ and is comprised between $-300$ E and $180$ E.

\subsection{Verticality}
In our setup, verticality can be ensured with an error of about $\delta\theta = 80\,\mu$rad.
As a consequence, supplementary terms in the phase shift arise from the projection of gravity on the horizontal axis which leads to a transverse velocity of the atoms $\delta v=g\delta\theta T$ and a sensitivity to rotation rate with respect to horizontal axis $\Omega_y$. According to \cite{Cadoret2016} this supplementary phase shift is expressed as $\phi=4k\Omega_y \delta v T^2=0.14$ mrad corresponding to an uncertainty of $\pm 67$ E on gravity gradient with $\Omega_y$ the Earth rotation rate.

\subsection{Drift of classical accelerometer}
The gravity gradient measurement using the correlation technique is sensitive to errors on the measurement of vibrations such as the bias $a_b$ of the classical accelerometer and its drift $\frac{da_{b}}{dt}$ during the measurement. We have estimated this drift by measuring the output signal of the accelerometer as a function of time during one day. The bias drift has been estimated from a linear fit to the data to  $\frac{da_{b}}{dt}=-0.47\times 10^{-9}\,g/s$. From this measurement we have calculated the bias phase induced by the bias drift of our mechanical accelerometer:
\begin{equation}
\varphi_b=4k_{\mathrm{eff}}\frac{da_{b}}{dt}T\Delta T \times T_{\Delta T}+2k_{\mathrm{eff}}\frac{da_{b}}{dt}T^3
\end{equation}
The first contribution to the bias phase arises from the change in acceleration bias between 2 measurements performed at $\Delta T$ and $-\Delta T$ respectively. In our experimental protocol,the time assymmetry is changed after a time $T_{\Delta T}=1$ hour leading to a bias phase $ \simeq 1.26$ mrad corresponding to a bias $\Delta \Gamma=900$ E.\newline
The second effect arises from the contribution of the drift during the interferometer integration time. This term gives rise to a bias on the gravity gradient measurement of 
$\Delta \Gamma=\frac{da_b/dt}{2gT}=62$ E.

\subsection{Effect of magnetic field}
Our interferometer sequence is applied to atoms selected in the $m_F=0$ state. Nevertheless, a quadratic Zeeman shift and an inhomogeneity in the bias magnetic field  applied to the atoms during the interferometer  induces an additional phase shift and a bias on the gravity gradient measurement that can be calculated using \cite{Storey94}. Thanks to our effective wave vector reversal protocol, this effect is mainly canceled and the associated bias  is $\Delta\Gamma=100$ E $\pm$ 100 E.

\subsection{Self attraction effect}
Because our experimental setup is not massless, one has to take into account the gravitationnal attraction of the upper part and lower part of the titanium vaccuum chamber on the atomic sensor.
The mass difference between the two parts is estimated to $\Delta m=1.5$ kg. Thus, using a point mass calculation approximation, one can estimate this mass difference to induce an artificial gravity gradient effect of the form $\Delta\Gamma=\frac{2G\Delta m}{r^3}$ where $G=6.67\times 10^{-11}$N.m$^2$kg$^{-2}$ is the gravitational constant and $r$ the distance between the masses. Assuming the distance from the atoms to be between 5 cm and 10 cm the effect is comprised between 1600 E and 200 E respectively.

\subsection{Conclusion on systematic effects}
The results of the main systematics are summarized in Table $\ref{tab:syst}$.
We finally obtained the gravity gradient values $\Gamma_{zz}^{(1)}=(6069\pm \,2459)\mathrm{E}$ and $\Gamma_{zz}^{(2)}=(5173\pm 3428 )\,\mathrm{E}$ which are statistically in agreement. For both methods the dominant systematic uncertainty comes from TPLS effect. The correlation technique remains limited by statistical uncertainty, the dominant noise coming from the technical detection noise.
\begin{table}
\caption{Correction (Corr.) and uncertainties (u) in E\"otvos [E] of the main systematic effects affecting the cold atom gradiometer using two different measurement mehods: (1) vibration isolated (2) Without vibration isolation.}
\begin{ruledtabular}
\label{tab:syst}
\begin{tabular}{ccccc}

			\rule[0cm]{0pt}{0.2cm} Source& Corr.  & u &Corr. & u\\
			\rule[0cm]{0pt}{0.2cm}  & (1)&(1)&(2)&(2)\\
			\hline
			\rule[0.2cm]{0pt}{0.2cm} Effect of slope&-4351&430&-& -\\
			\rule[0.2cm]{0pt}{0.2cm} Two-photon light shift& $3920$ & $2019$ &3920&2019\\
			\rule[0.2cm]{0pt}{0.2cm} Sensitivity to initial velocity& $0$ & $300$ &0&300\\
\rule[0.2cm]{0pt}{0.2cm} Additional laser line&0 & [-310;180]&0&[-310;180]\\
			\rule[0.2cm]{0pt}{0.2cm} Verticality&0&67&0&67 \\
			
						\rule[0.2cm]{0pt}{0.2cm} Accelerometer drift&- &-&-1.5&962\\
									\rule[0.2cm]{0pt}{0.2cm} Effect of magnetic field&-100&100&-100&100\\
									\rule[0.2cm]{0pt}{0.2cm} Self attraction effect&-1000&1000&-1000&1000\\
								\rule[0.2cm]{0pt}{0.2cm} Statistical u.&-&766&-&2355\\
									\hline
									\rule[0.2cm]{0pt}{0.2cm} Total&-1531&$2459$&2818&3428\\

\end{tabular}
\end{ruledtabular}
\end{table}



\section{Discussion and possible improvements}
\label{Discussion}

We have compared the scale factor $\left(\mathcal{S}=\partial \Delta\Phi/\partial \Gamma\right)$ of our four pulse AI with the one of a conventional atomic gradiometer using two pairs of Mach-Zehhnder-like atom accelerometers in differential mode. For this comparison we have adressed both Earth-based, and space-based (e.g microgravity) issues. The calculation assumes a total interrogation time $T_{\mathrm{int}}$ and apparatus length $L$. We have neglected the recoil effect term. The recoil term is responsible of a displacement of the atoms that we have assumed to be small compared to the length of the apparatus for our interrogation time. Nevertheless, this term should be considered for long interaction time where the atom displacement is larger. Results are presented  in table \ref{tab:tab3}. From \ref{tab:tab3}, in a microgravity environment, a dual Mach-Zehnder atom accelerometer geometry has a scale factor 8 times larger than the one of a four-pulse AI. However, using a single cloud atomic gradiometer for terrestrial applications could be of interest as the reduction in its scale factor is only a factor of 2. This scale factor reduction has  to be  put in balance  with the ease of implementation of such an experimental setup which only requires \emph{(i)} one atomic source, \emph{(ii)} an additional Raman $\pi$-pulse with respect to a conventional Mach-Zehnder atomic gravimeter.\newline
One can calculate the gravity gradient sensitivity $\delta \Gamma$ that can be obtained using our technique and considering an Earth based gravity gradient measurement with an interrogation time $T_{int}=4T$ corresponding to a 1 meter length apparatus. We have taken the case where the atoms are not launched.
Assuming that the ultimate phase resolution is quantum projection noise limited $\delta(\Delta\Phi)\approx 1/\sqrt{N_{at}}$ the single-shot gradiometer sensitivity is given by:
\begin{equation}
\delta\Gamma \approx \frac{2}{6 C\sqrt{N_{at}}k_{\mathrm{eff}}gT^4}
\end{equation}

Assuming $10^6$ detected atoms, a total interrogation time of $4T\simeq 452$ ms, a contrast of $C=0.5$ wich can be obtained with a thermal atomic sample using adiabatic passage technique \cite{Jaffe2018} one obtains $\delta\Gamma\approx 13$ E/$\sqrt{Hz}$.
The contribution of the mechanical accelerometer self-noise to the sensor's sensitivity is estimated to be $\approx 1$ E/$\sqrt{Hz}$.
\newline
We can expect an improvement of our atomic gradiometer. When using the correlation technique, an increase in sensitivity could be obtained \emph{(i)} by changing the sign of the timing assymetry $\Delta T$ at the repetition rate of the experiment, hence reducing the effect of the classical accelerometer drift by a factor of 3600,\emph{(ii)} replacing the fringe fitting procedure by a more sensitive FLM technique  using the signal of the classical accelerometer to compensate in real time the phase shift of the AI as in \cite{Lautier2014}. When measuring the gravity gradient using the vibration isolation platform,  applying a  phase step $\delta\varphi$ on the Raman laser phase rather than changing the chirp rate $\alpha$ when performing the FLM technique, would allow to cancel the slope effect as the same slope would appear for each of the four fringe patterns.\newline
In the scope of field applications, using a single-cloud double-loop AI geometry allows to be insensitive to Coriolis force \cite{Marzlin1996} which is responsible of both a bias  and mostly a severe loss of contrast when performing the AI on a  boat \cite{Bidel2018} or a plane \cite{Geiger2011} if no rotation compensation is used such as a tip-tilt mirror \cite{Lan2012} or a gyrostabilized platform.

\begin{table}[h]
\caption{Comparison in scale factors $\mathcal{S}$, of Earth-based dual cloud atomic gradiometer with respect to a single-cloud gradiometer. $L$:apparatus length; $g$:gravity acceleration; $T_{int}$:total interrogation time. $T_{int}=2T$ ($4T$) for a Mach-Zehnder type geometry and Four-pulse geometry respectively.}
\begin{ruledtabular}
\label{tab:tab3}
\begin{tabular}{cccc}

			\rule[0cm]{0pt}{0.2cm} Environment & Mach-Zehnder&4-pulse AI& Scale factor ratio\\
			\rule[0cm]{0pt}{0.2cm}   &$\mathcal{S}_{MZ}$&$\mathcal{S}_{4P}$ &$\mathcal{S}_{MZ}/\mathcal{S}_{4P}$\\
			\hline
			\rule[0.2cm]{0pt}{0.2cm} Microgravity&$\frac{1}{4}kLT_{int}^2$&$\frac{1}{32}kLT_{int}^2$&8 \\
			\rule[0.2cm]{0pt}{0.2cm} Earth-based&$\frac{1}{8}\frac{kL^2}{g}$&$\frac{1}{16}\frac{kL^2}{g}$&2\\	
\rule[0.2cm]{0pt}{0.2cm} Earth-based&  &&\\
\rule[0.2cm]{0pt}{0.2cm} + atom launch&$\frac{1}{4}\frac{kL^2}{g}$ &$\frac{27}{256}\frac{kL^2}{g}$&$\approx 2.37$\\	
\end{tabular}
\end{ruledtabular}
\end{table}

\section{Conclusion}
\label{Conclusion}
In conclusion, we have demonstrated an experimental proof-of-principle measurement of the vertical gravity gradient using a single proof mass of cold $^{87}$Rb atoms and a four-pulse double-loop AI. We performed the measurements first using a vibration isolator and then demonstrating the correlation technique in presence of vibration noise. The results obtained using both methods are in fair agreement  despite a reduced sensitivity in comparison with state-of-the-art atomic gradiometer \cite{Sorrentino2014}, due to a rather short interrogation time ($4T=154.4$ ms). 
Better performances of the four-pulse gradiometer are expected with larger interrogation time and using efficient atom optics techniques such as  adiabatic rapid passage optical pulses \cite{Jaffe2018} which does not require a colder atomic sample.
Finally, the results obtained in a strap down configuration using the correlation technique, combined with the simplicity of our experimental setup, (single source and unique detection), the ability to switch easily from a gravimeter to a gradiometer sensor immune to constant rotation, could be of interest for the realization of  instruments dedicated to gravity or gradiometry measurements in noisy environments.

\begin{acknowledgments}
We thank F. Nez, from Laboratoire Kastler Brossel (LKB), for his help on the project.

\end{acknowledgments}

\appendix





\bibliography{bibliography_of_article2}

\end{document}